\documentclass[11pt]{article}

\usepackage{amssymb} %
\usepackage{graphicx,amsmath,amsfonts}
\usepackage{color}
\usepackage[compatibility=false]{caption}
\usepackage{subcaption}
\usepackage{comment}
\usepackage{nicefrac}
\usepackage{mathrsfs}
\begin{document}

\newcommand{\naturals}{{{\mathbb{N}}}}
\newcommand{\integers}{{{\mathbb{Z}}}}
\newcommand{\rationals}{{{\mathbb{Q}}}}
\newcommand{\reals}{{{\mathbb{R}}}}

\newcommand{\cc}{{{{\mathrm{CC}}}}}
\newcommand{\monroe}{{{{\mathrm{Monroe}}}}}
\newcommand{\stv}{{{{\mathrm{STV}}}}}
\newcommand{\sntv}{{{{\mathrm{SNTV}}}}}
\newcommand{\bloc}{{{{\mathrm{Bloc}}}}}
\newcommand{\kborda}{{{{k\hbox{-}\mathrm{Borda}}}}}
\newcommand{\rep}{{{\mathrm{rep}}}}
\newcommand{\capacity}{{{{\mathrm{cap}}}}}
\newcommand{\cost}{{{{\mathrm{cost}}}}}
\newcommand{\pos}{{{{\mathrm{pos}}}}}
\newcommand{\argmax}{{{{\mathrm{argmax}}}}}
\newcommand{\argmin}{{{{\mathrm{argmin}}}}}
\newcommand{\borda}{{{{\mathrm{B}}}}}
\newcommand{\bordainc}{{{{\mathrm{B, inc}}}}}
\newcommand{\bordadec}{{{{\mathrm{B, dec}}}}}
\newcommand{\OPT}{{{{\mathrm{OPT}}}}}
\newcommand{\opt}{{{{\mathrm{opt}}}}}
\newcommand{\sat}{{{{\mathrm{sat}}}}}
\newcommand{\low}{{{{\mathrm{low}}}}} 
\newcommand{\high}{{{{\mathrm{high}}}}} 
\newcommand{\E}{\mathop{\mathbb E}} \newcommand{\w}{{{{\mathrm{w}}}}}

\newcommand{\NN}{\mbox{$\cal N$}}
\newcommand{\PP}{\mbox{$\cal P$}}
\newcommand{\WW}{\mbox{$\cal W$}}
\newcommand{\RR}{\mbox{$\cal R$}}
\newcommand{\A}{\mbox{$\cal A$}}
\newcommand{\calS}{\mbox{$\cal S$}}
\newcommand{\poly}{\mathrm{poly}}

\def\R{\mathbb{R}}
\def\Z{\mathbb{Z}}
\def\N{\mathbb{N}}
\def\ds{\displaystyle}
\def\oa#1{\overrightarrow{#1}}
\def\ola#1{\overleftarrow{#1}}
\def\row#1#2{{#1}_1,\ldots ,{#1}_{#2}}
\def\brow#1#2{{\bf {#1}}_1,\ldots ,{\bf {#1}}_{#2}}
\def\rrow#1#2{{#1}_0,{#1}_1,\ldots ,{#1}_{#2}}
\def\urow#1#2{{#1}^1,\ldots ,{#1}^{#2}}
\def\irow#1#2{{#1}_1,\ldots ,{#1}_{#2},\ldots}
\def\lcomb#1#2#3{{#1}_1{#2}_1+{#1}_2{#2}_2+\cdots +{#1}_{#3}{#2}_{#3}}
\def\blcomb#1#2#3{{#1}_1{\bf {#2}}_1+\cdots +{#1}_{#3}{\bf {#2}}_{#3}}
\def\2vec#1#2{\left(\begin{array}{c}{#1}\\{#2}\end{array}\right)}
\def\threevec#1#2#3{\left[\begin{array}{r}{#1}\\{#2}\\{#3}\end{array}\right]}
\def\mod#1{\ \hbox{\rm (mod $#1$)}}
\def\gcd#1#2{\hbox{gcd}\>(#1,#2)}
\def\lcm#1#2{\hbox{lcm}\>(#1,#2)}
\def\card#1{\hbox{card}\>(#1)}
\def\adv{\hbox{adv}}
\def\union{\cup}
\def\intsct{\cap}
\def\Union{\bigcup}
\def\Intsct{\bigcap}

\newcommand{\calA}{\mathcal{A}}
\newcommand{\calE}{\mathcal{E}}
\newcommand{\calR}{\mathcal{R}}
\newcommand{\calF}{\mathcal{F}}
\newcommand{\calB}{\mathcal{B}}
\newcommand{\calC}{\mathcal{C}}
\newcommand{\calQ}{\mathcal{Q}}
\newcommand{\calD}{\mathcal{D}}
\newcommand{\calP}{\mathcal{P}}
\newcommand{\Dec}{\mathit{Dec}}
\newcommand{\revnot}[1]{\overleftarrow{#1}}

\newcommand{\peak}{{\mathrm{peak}}}
\newcommand{\dfs}{{\mathrm{dfs}}}

\newcommand{\p}{{{\mathrm{P}}}}
\newcommand{\np}{{{\mathrm{NP}}}}
\newcommand\mydots{\hbox to 1em{.\hss.\hss.}}
\sloppy

\title{Modeling Representation of Minorities Under Multiwinner Voting Rules\\
(extended abstract, work in progress)}

\author{Piotr Faliszewski\\ 
        AGH University\\
        Poland\\
        \phantom{Weizmann Institute of Science}
  \and
        Jean-Fran\c{c}ois Laslier\\
        Paris School of Economics\\
        France\\
        \phantom{Weizmann Institute of Science}
  \and
        Robert Scheafer\\
        AGH University\\
        Poland\\
        \phantom{Weizmann Institute of Science}
  \and
       Piotr Skowron\\
       Oxford University\\
       United Kingdom\\
        \phantom{Weizmann Institute of Science}
 \and
       Arkadii Slinko\\
       University of Auckland\\ 
       New Zealand\\
        \phantom{Weizmann Institute of Science}
 \and
       Nimrod Talmon\\
       Weizmann Institute of Science\\
       Israel
}

\maketitle

\begin{abstract}
  The goal of this paper is twofold.  First and foremost, we aim to
  experimentally and quantitatively show that the choice of a
  multiwinner voting rule can play a crucial role on the way
  minorities are represented.  
 We also test the possibility for some of these rules to achieve proportional representation.
\end{abstract}

\section{Introduction}
The use of voting rules as a mean of manipulation to advantage or
disadvantage minorities is widespread.  With the passage of the Voting
Rights Act in 1965 in the United States, the right of minorities to
register and vote was largely secured. It was soon discovered,
however, that minority voting did not guarantee the election of
minorities or minority-preferred candidates. This was a result of a
widespread use of manipulation by the choice of voting
rules~\cite{grofman1992controversies,grofman1994minority,trebbi2008electoral}.
Manipulation of electoral rules, however, is not a
prerogative exclusive of American cities. Pande~\cite{pande2003can}
provides a discussion of electoral rules and racial politics in
elections in India. Alexander~\cite[p.~211]{alexander2004france}
describes in detail the 1947 Gaullist manipulations of electoral rules
in France; in the Paris area, where the Gaullist alliance was weak,
they introduced proportional representation but in rural areas, where
the alliance was strong, they introduced plurality.
Kreuzer~\cite[p.~229]{kreuzer2004germany} describes strategic
manipulation of voting rules in postwar Germany.

In this paper we undertake an experimental study of the effect that
some voting rules have on representation of minorities.  
The American literature has dealt at length with manipulation by
re-districting, often called ``gerrymandering,'' that is
crafting the electoral districts to the advantage of the
designer~\cite{grofman1982representation}.  
In the present paper, we do not tackle the districting question. 
Our work applies to
the case of a district that elects $k>1$ delegates as well as to the,
formally equivalent, case of a country that does not uses districting
for electing its Parliament. Moreover, we consider the rules which
take into account not only the first preferences of voters but also
the second, third and further preferences. 
For these rules not based on districting, the aspects of
the causal connection between electoral systems and vote-seat
disproportionality remains obscure~\cite{powell2000election}.

We adopt a standard spatial two-dimensional model of voting, assuming that both
voters and candidates have ideal political positions on the plane and Euclidean preferences.
Applied research has shown that having two dimensions is often
sufficient to have meaningful descriptions of voters' political
opinions \cite{schofield2007spatial}.  The idea for this paper stems
from a previous work of Faliszewski, Sawicki, Schaefer and Smo{\l}ka
regarding a selection method for genetic algorithms based on
multiwinner voting~\cite{fal-saw-sch-smo:c:multiwinner-genetic-algorithms}.

\section{Preliminaries}

\paragraph{Elections and Voting Rules}
Let $V = \{v_1, \ldots, v_n\}$ be the set of $n$ \emph{voters} and $C
= \{c_1, \ldots, c_m\}$ be the set of $m$ \emph{candidates}.  The
voters have their intrinsic preferences over candidates, which are
represented as preference orders (i.e., rankings of the candidates
from best to worst).  By $\pos_v(c)$ we denote the position of
candidate $c$ in the preference ranking of voter $v$.  For example, a
voter $v$ who likes $c_1$ best, then $c_2$, then $c_3$, and so on,
would have preference order $c_1 \succ c_2 \succ \cdots \succ
c_m$. For this voter, we would have $\pos_v(c_1) = 1$, $\pos_v(c_2) =
2$, and so on.

We are interested in multiwinner elections, where the goal is to
select a committee of size $k$ (i.e., a size-$k$ subset of $C$).  A
\emph{multiwinner election rule} is a formal decision process that
given preferences of the voters and a positive integer $k \in
\naturals$ returns a committee that, according to this rule, is most
preferred by the population of the voters viewed as a whole.

Many multiwinner rules rely on the notion of score for the candidates.
For each integer $t \in \{1,\ldots,m\}$, the $t$-Approval score of candidate $c$
in vote $v$ is $1$ if $v$ ranks $c$ among top $t$ positions, and is
$0$ otherwise. The Borda score of candidate $c$ in vote $v$, denoted
$\beta(c,v)$, is $m - \pos_v(c)$. The Plurality score of a candidate
is his or her $1$-Approval score. Given one of these notions of score,
the total score of a candidate in the election is the sum of his or
her scores from all the voters.

The following rules are considered in this paper:

\begin{description}
\item[Single Nontransferable Vote (SNTV).] SNTV selects a committee
  that consists of those $k$ candidates with the highest Plurality
  scores.\smallskip

\item[Bloc.] Bloc selects a committee that consists of those $k$
  candidates with the highest $k$-Approval scores (one can think of
  Bloc as if each voter gave a point to each candidate from his or her
  ideal committee).\smallskip

\item[$\boldsymbol{k}$-Borda.]  $k$-Borda selects a committee that
  consists of those $k$ candidates with the highest Borda scores.  In
  the world of single-winner voting rules ($k=1$), Borda is usually
  seen as electing some kind of compromise candidate.\smallskip

\item[Chamberlin--Courant Rule.] For each voter $v$ and each committee
  $C$ a \emph{representative of $v$ in $C$} %
  is the most preferred member of $C$, according to $v$. The
  Chamberlin--Courant rule~\cite{cha-cou:j:cc} selects a committee so
  that the sum of the Borda scores of the voter representatives is
  maximized (alternatively, one can think of minimizing the average
  position of a voter's representative). Formally, the
  Chamberlin--Courant rule selects a committee $C$ that maximizes the
  value $\sum_{v \in V} (\max_{c\in C}\beta(c,v))$.
  Unfortunately, computing a winning committee under the
  Chamberlin--Courant rule is
  $\np$-hard~\cite{pro-ros-zoh:j:proportional-representation,bou-lu:c:chamberlin-courant}. For
  the purpose of this paper, we were able to compute
  Chamberlin--Courant results using its formulation as an integer
  linear program (ILP) by running the CPLEX optimization package.  Lu
  and Boutilier~\cite{bou-lu:c:chamberlin-courant} and Skowron et
  al.~\cite{sko-fal-sli:j:multiwinner} offer %
  approximation
  algorithms that one could use for larger elections.  \smallskip

\item[Monroe Rule.] Monroe~\cite{mon:j:monroe}, similarly to
  Chamberlin and Courant, explored the concept of a representative of
  a voter. He, however, required that each committee member should
  represent roughly the same number of voters. A function $\Phi\colon
  V \to A$ is a Monroe assignment for a committee $C$ if for each
  candidate $a \in C$ it holds that $\left\lfloor \nicefrac{n}{k}
  \right\rfloor \leq \Phi^{-1}(a) \leq \left\lceil \nicefrac{n}{k}
  \right\rceil$. Intuitively, Monroe assignments represent valid
  mappings between the voters and their representatives. Let
  $\mathscr{A}(C)$ denote the set of all Monroe assignments for a
  committee $C$. According to the Monroe rule, the score of committee
  $C$ is defined as $\mathrm{score_M}(C) = \max_{\Phi \in
    \mathscr{A}(C)}(\sum_{v \in V}\beta(\Phi(v),v))$. The committee
  $C$ that maximizes $\mathrm{score_M}(C)$ is selected as the winner.
  Intuitively speaking, the idea behind the Monroe rule is to
  partition the electorate into roughly same-sized
  districts\footnote{Note that these ``virtual districts'' are based
    on voters' preferences and not on geographical location.} and
  assign to each district a distinct candidate with as high Borda
  score as possible.  Just like the Chamberlin--Courant rule, Monroe
  rule is $\np$-hard to
  compute~\cite{pro-ros-zoh:j:proportional-representation}, but this
  time for most of our experiments the ILP formulation turned out to
  be too complex for CPLEX to solve within reasonable amount of
  time. Thus, instead we used the Greedy-Monroe approximation
  algorithm of Skowron et
  al.~\cite[Algorithm~A]{sko-fal-sli:j:multiwinner} which is
  guaranteed to select a committee $C$ whose $\mathrm{score_M}(C)$ is
  close to being the maximum.\smallskip

\item[Single Transferable Vote (STV).] STV is a multi-round procedure
  that operates as follows. In each round, if there exists a candidate
  $c$ who is preferred the most by at least $q = \left\lfloor
    \nicefrac{n}{k+1} \right\rfloor + 1$ voters, then $c$ is added to
  the winning committee. At the same time we remove from further
  consideration exactly $q$ voters which rank $c$ on top, and delete
  $c$ from the preference rankings of all other voters. Otherwise,
  i.e., if each candidate is most preferred by less than $q$ voters,
  then we select a candidate which is most preferred by the smallest
  number of voters and delete this candidate from preference rankings
  of all voters.\footnote{Occasionally, we run into trouble when
    computing STV winners. For example, for $n = 600$ voters and
    committee size $k = 52$ we should use quota value $q = \lceil(
    \nicefrac{600}{53} \rceil + 1 = 12$. In each round in which STV
    puts a candidate into a committee, it also deletes $q$
    voters. Yet, $k \cdot q = 624$ so we do not have enough
    voters. Fortunately, in our experiments such situations were
    occurring only for committee sizes over 50. Thus we do not give
    results for STV for committees of sizes larger than 50.}  We note
  that this description of STV is not complete and there is a lot of
  room for various tie-breaking decisions (for example, it is not
  obvious which voters exactly to delete when a candidate is added to
  the committee). We describe our approach to tie-breaking below.  
  See Tideman and
  Richardson~\cite{tid-ric:j:stv} for an overview of the STV rule and
  its variants.
\end{description}

\noindent The next two rules do not exactly fit in our framework because they are based on districting. 

\begin{description}
\item[First Past The Post (FPP).] Under FPP voters are divided into
territorial districts (constituencies) of approximately equal sizes
and each constituency elects their own representative by using the
Plurality rule (i.e., the candidate with the highest Plurality score
wins within the constituency).  \smallskip

\item[District-Based Borda.] The same as FPP, but with the use of
Borda scores instead of Plurality scores. 
\end{description}

We shall consider these two last rules under the assumption of random
districting. This means that we assume that any territorial district
represents an unbiased collection of the political opinions, and we
create ``districts'' artificially by choosing a random partition of
the electorate. We thus obtain two voting rules that could be called
``Random district FPP'' and ``Random district Borda.'' These rules
deserve to be studied as benchmarks for comparison with the others.

Occasionally, our voting rules run into situations where they have to
break ties (this is particularly imminent in the definition of STV,
but all rules face this issue). To simplify our experiments, we break
all ties, whenever they occur, uniformly at random.

%
%
%
%
%
%
%
%
%
%
%
%
%
%
%
%
%

%

%
%
%
%
%
%
%
%
%
%
%
%
%
%
%
%
%
%
%
%
%
%
%
%
%
%
%
%
%
%
%
%
%
%

%
%
%
%
%
%
%
%
%
%
%
%
%
%
%
%
%
%
%
%
%
%
%
%
%
%
%
%
%
%
%
%

%
%
%
%
%
%

%
%
%

%
%
%
%
%
%
%
%
%
%

%
%
%
%
%
%
%
%
%
%
%

%
%
%
%
%
%
%
%
%
%
%
%
%
%
%
%
%
%
%
%
%
%
%
%
%

%
%
%
%
%
%
%
%
%

%
%
%
%
%

%
%
%
%
%
%
%
%
%
%
%
%
%
%
%
%
%
%
%
%

%
%
%
%
%
%
%
%
%
%
%
  
%
%
%
%
%
%
%
%
%
%
%

%
%
%
%
%
%

%
%
%
%
%
%
%
%
%
%
%

%
%
%
%
%
%
%
%
%
%
   
%

%
%
%
%
%
%
%
%
%
%
%
%
%
%
%

%
%
%
%
%
%
%
%
%
%
%
%
%
%
%
%
%
%

%
%
%
%
%
%
%
%
%
%
%
%
%
%
%
%
%
%

%
%
%
%
%
%
%

%
%
%
%

\paragraph{Spatial Models of Elections}
Euclidean preferences \cite{DavisHinich66} stipulate that both candidates
and voters can be represented as points in an Euclidean space, and
that voters rank candidates according to the increasing order of
Euclidean distances from themselves. The idea is that points
correspond to political programs. Candidates are represented by their
actual programs, whereas voters are represented by the ideal programs
they believe in \cite{Plott1967,mckelvey1990,enelow1984spatial}.  

As the empirical analysis of elections
shows~\cite{schofield2007spatial}, the dimension of the political
space seldom exceeds two. Usually, the left-right spectrum is the main
one and the second dimension could be, for example, caused by the
influences of religion. In our model we assume that voters
have two-dimensional Euclidean preferences.

\section{Results}

We present results of two experiments. The purpose of the first experiment is to get an initial understanding of the rules
discussed. The purpose of the second one is to asses how these
rules treat minorities. 

\subsection{Initial Experiment: On Representativity}

The voting rule in a representative democracy ideally accomplishes two
tasks: selects a representative set of delegates (e.g., a parliament)
and assigns voters to delegates. This means the two main purposes of
a voting rule is to achieve a certain level of representativity and a
certain level of accountability. These two requirements are not easy
to combine. One standard solution to this is to use
First-Past-the-Post (FPP), a system which operates with electoral
(usually territorial) districts of approximately the same size and
allows voters in each district to elect their representative using
Plurality. This perfectly solves the problem of accountability but the
representativity of such a system is known to be poor because it tends
to be detrimental for minorities, especially for a minority that is
spread in all districts. On the other hand, party-list
proportional-representation systems~\cite{puk:b:pr} can be quite good
on representativity, provided that the threshold of representation is
small, but very poor on accountability. 

There seems to be a certain tension between accountability and
representativity of multiwinner voting rules as well, and some rules
seem to accommodate both desires better than others.
While we do not yet have a good measure of voting rules'
accountability, in this section we attempt to evaluate the
representativity of their outcomes.  Our idea is simply that a rule is
more representative when it is more likely for each voter that some
candidate with similar political views is elected.

\paragraph{Misrepresentation}  Formally, we take the following
approach. Let $d$ denote the Euclidean distance in our two-dimensional
space of political programs. Given a voter $v$ and a winning committee
$W$, we define $\Psi(v)=\min_{c\in W} d(v,c)$ to be the distance
between $v$ and the closest member of $W$.  If we view distances as
meaningful characteristics of preferences, it is natural to consider
$\Psi(v)$ as a measure of $v$'s misrepresentation in the
committee. For an election $E = (C,V)$ and a committee $W$, we define
$D(E,W) = \frac{1}{|V|}\sum_{v \in V}\Psi(v)$ to be the average
misrepresentation of the voters.

Note that our definition does not embody any notion of efficiency. As
an example, imagine that a small group of voters is very homogeneous
and has preferences very different from the rest of the electorate. If
this group elects a single delegate, representation can be very good
for this group, according to our definition. But, depending on how the
decisions are taken in the Parliament, it may well be that this
delegate has no real power. 

\paragraph{Candidates}  Of course representativity chiefly
depends on who are the candidates. %
To focus on the effect of
the voting rule itself, we consider in this paper that the set of
candidates, on its own, is a good representation of the
electorate. This is easily done by drawing candidates' political
platforms from the same distribution as the voters ideal points. At
least for large values of $k$, this achieves the goal.\footnote{The
  assumption that the set of candidates is identical to the set of
  voters is often met in the Political Economy literature since
\cite{osborne1996model,besley1997economic} and labelled the
  ``citizen-candidate'' model.} 

\begin{figure}
  \centering
  \includegraphics[height=8cm]{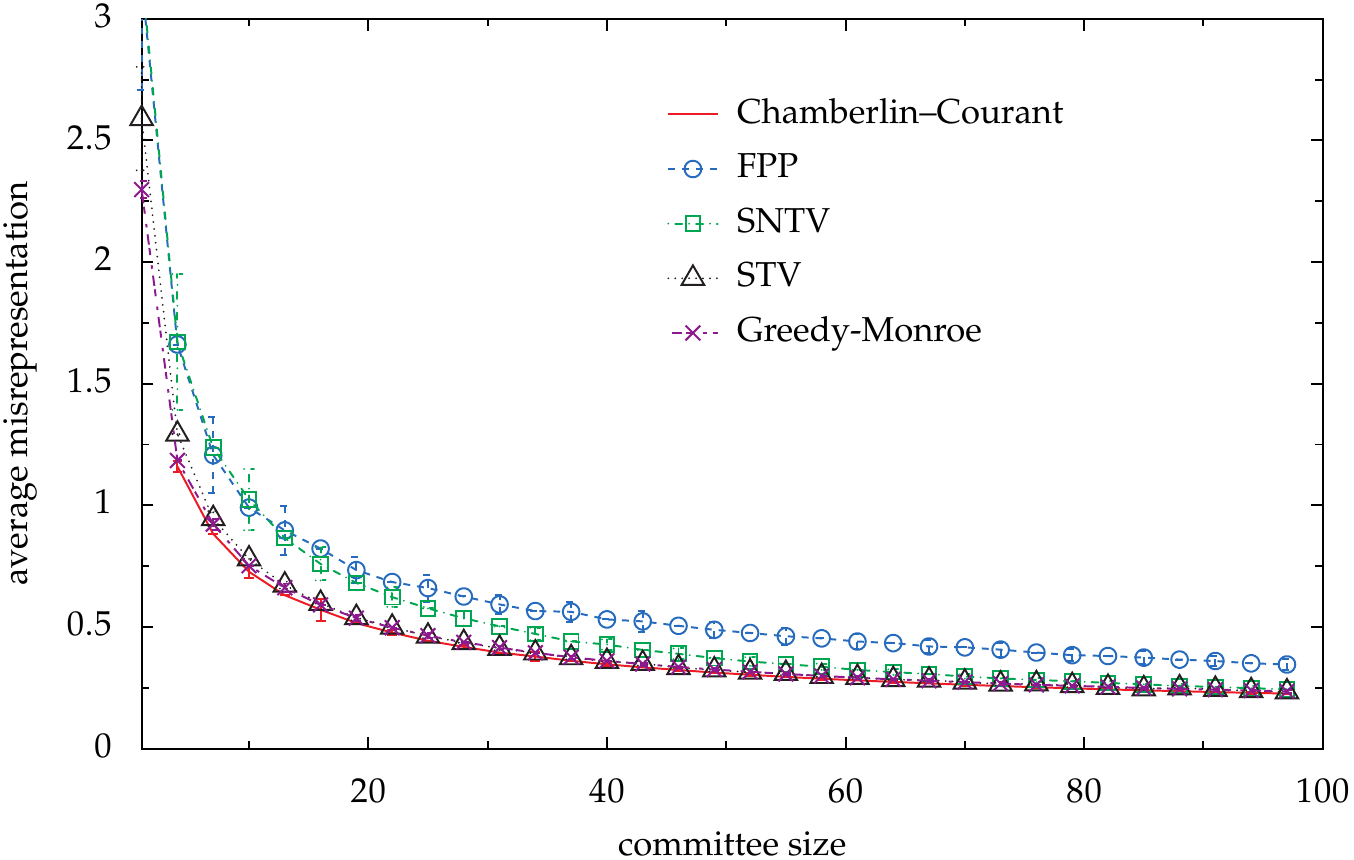}
  \caption{\label{fig:satisifaction1}Average
    misrepresentation of the voters for rules that aim at achieving
    proportional representation. The vertical bars indicate standard
    deviation.}

\medskip

  \centering
  \includegraphics[height=8cm]{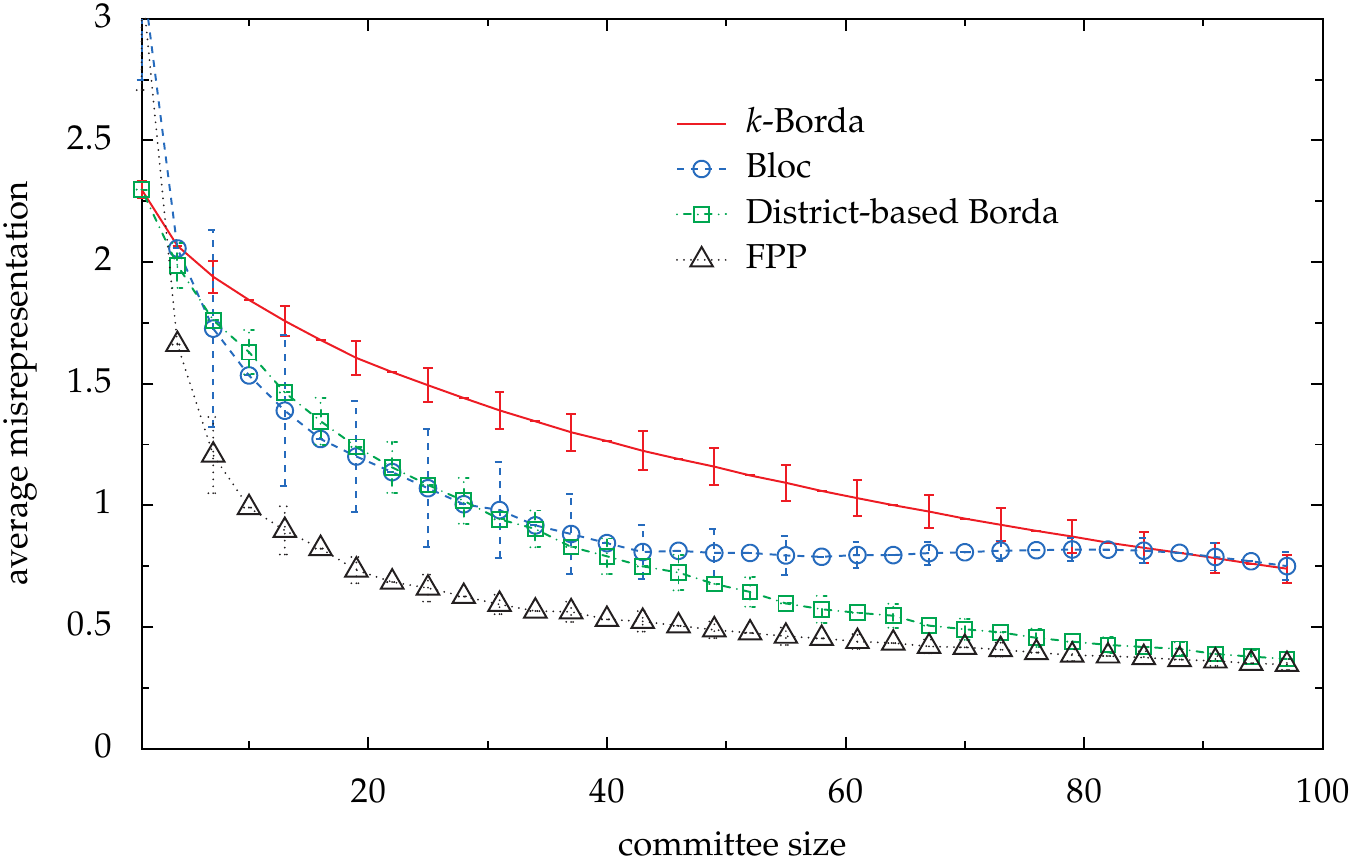}
  \caption{\label{fig:satisifaction2}Average misrepresentation of the
    voters for the other rules. The vertical bars indicate standard
    deviation.}
\end{figure}

\paragraph{Results} We have measured the average
misrepresentation for our rules in the following setting.  We 
generated $60$ elections with $300$ candidates and $600$ voters each,
all distributed uniformly on a $6 \times 6$ square. For each election
we have computed the results of all our voting rules, for committees
of sizes from $1$ to $97$ with a step of $3$. For each case we have
computed the average misrepresentation of the voters. We present our
results on Figures~\ref{fig:satisifaction1}
and~\ref{fig:satisifaction2}.
Absolute values of the computed average
misrepresentation is not very meaningful
, and thus one should focus on relative comparison of the voting rules.

On Figure~\ref{fig:satisifaction1} we show the results for
Random-District-FPP, SNTV, STV, Greedy-Monroe, and
Chamberlin--Courant.
We can see that STV, Greedy-Monroe, and Chamberlin--Courant achieve
next to indistinguishable results. SNTV achieves somewhat worse
results (but for large committees it converges with the previous
three), and FPP does not converge to the others even for very large
committees.

On Figure~\ref{fig:satisifaction2} we show the result for Bloc,
$k$-Borda, Random-District-Borda, and FPP. $k$-Borda is the least
proportional rule (indeed, inspection of the results has shown that
$k$-Borda picks a cluster of candidates in the center of our square;
it is designed to find candidates that are least offensive to all the
voters).  While adding random districts to Borda (i.e., considering
Random-District-Borda) helps significantly, the results are still
worse than for the rules from Figure~\ref{fig:satisifaction1}.  Bloc
also does poorly with respect to proportionality (it finds
concentration areas with many voters and chooses clusters of candidates
there; for large committees it tends to return the same or
similar committees as $k$-Borda).

There is a simple but important conclusions from this experiment.  For
the case of uniform distribution of candidates and voters, there seems
to be a single natural notion of 
representation of the
voters, and all our voting rules that were designed to find
correct representation (in the context of preferences orders)
appear to find it. It is quite remarkable since the definitions of our
rules can be significantly different (it certainly is not obvious that
STV and Chamberlin--Courant would be finding, in essence, the same
kinds of results). 

\subsection{A Polarized Society}

The choice of an electoral system has a major impact on the survival
of small political parties. The Liberal Democrats in the United Kingdom is
an example of such a party. They have some left-wing and some
right-wing policies so many researchers place them squarely in the
middle of the UK political spectrum. However, the existence of a
centrist party under FPP is extremely challenging\footnote{"Why being
  centrist hasn't helped the Lib Dems". New Statesman. 6 October
  2014. Retrieved 26 April 2016.}.  Even under the mixed-member
proportional (MMP) electoral system of New Zealand, centrist parties
often struggle, as exemplified by the virtual demise of Peter Dunne's
United Future party in 2013.

Here we deal with multiwinner voting rules that do not rest on the
existence of political parties. In order to explore the question of the
``squeezing of the center'' in this framework, we consider the
following situation.

The population itself is polarized in the sense that most voters are
extreme. Precisely, we suppose that the electorate is made of three
sub-populations: two large groups and a small one, with the small
group, the ``centrist voters'' in between the two large groups.  The
voters depicted by the black dots are taken from three Gaussian
distributions. The mean values for these Gaussians are, respectively,
(-2,0), (0,0), and (2,0); standard deviation is 0.25 in each case.
For the left and the right party, we generated 100 voters for each,
while for the centrist party we generated 50 voters (i.e., altogether,
there are 250 voters; the large parties have 40\% of the electorate
each, whereas the centrist party has 20\% of the electorate).

As to the candidates, we now suppose that they are not taken from the
same distribution as the voters, as in the previous experiment, but
that they are spread uniformly over the whole political spectrum
(there are 600 of them; depicted as gray points).
This leaves open the possibility to elect ``compromise'' candidates
that would lie in between two groups.

\begin{figure}[p!]
    \centering
    \begin{subfigure}{7cm}
      \centering
      \caption{\label{fig:kBorda}Results for SNTV}
      \includegraphics[height=3.25cm]{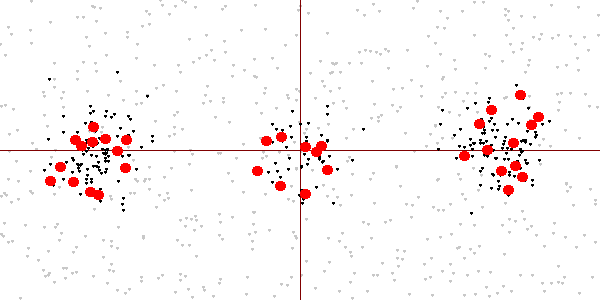}  
    \end{subfigure}
    \begin{subfigure}{7cm}
      \centering
      \caption{\label{fig:Bloc}Results for STV}
      \includegraphics[height=3.25cm]{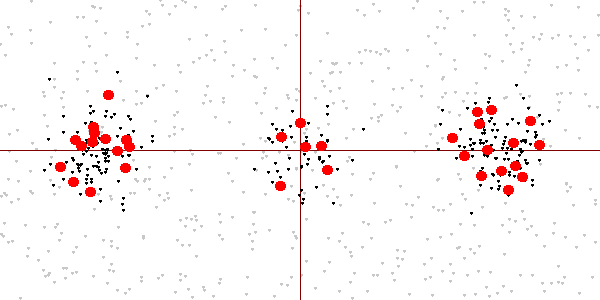}
    \end{subfigure}

    \bigskip\bigskip

    \begin{subfigure}{7cm}
      \centering
      \caption{\label{fig:kBorda}Results for Chamberlin--Courant}
      \includegraphics[height=3.25cm]{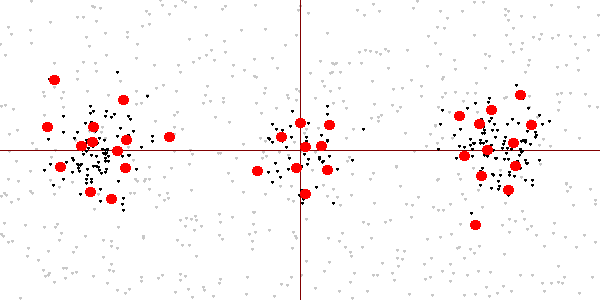}  
    \end{subfigure}
    \begin{subfigure}{7cm}
      \centering
      \caption{\label{fig:Bloc}Results for Greedy-Monroe}
      \includegraphics[height=3.25cm]{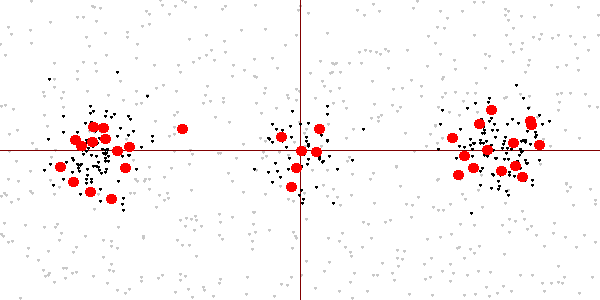}
    \end{subfigure}

    \bigskip\bigskip

    \begin{subfigure}{7cm}
      \centering
      \caption{\label{fig:kBorda}Results for FPP}
      \includegraphics[height=3.25cm]{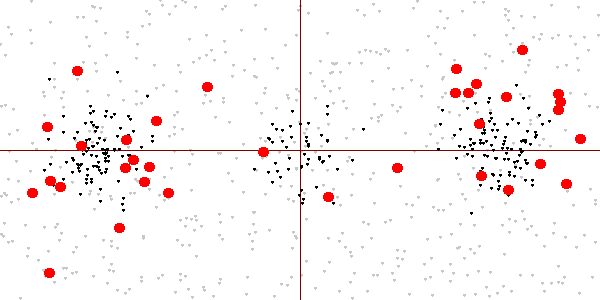}  
    \end{subfigure}
    \begin{subfigure}{7cm}
      \centering
      \caption{\label{fig:Bloc}Results for District-Based Borda}
      \includegraphics[height=3.25cm]{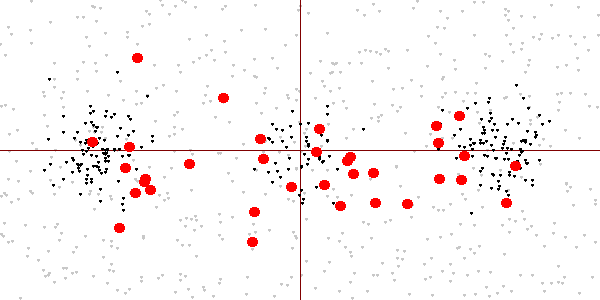}
    \end{subfigure}

    \bigskip\bigskip

    \begin{subfigure}{7cm}
      \centering
      \caption{\label{fig:kBorda}Results for Borda}
      \includegraphics[height=3.25cm]{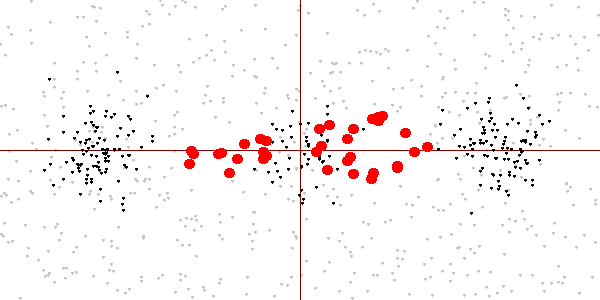}  
    \end{subfigure}
    \begin{subfigure}{7cm}
      \centering
      \caption{\label{fig:Bloc}Results for Bloc}
      \includegraphics[height=3.25cm]{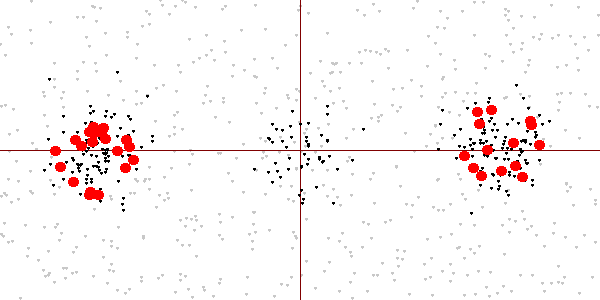}
    \end{subfigure}

    \caption{\label{fig:uni}Results for two big groups of voters and a smaller
      centrist one, for committee size $k=34$, for the case where
      600 candidates are distributed uniformly over the $6 \times 3$
      rectangle over the positions of the voters.}
\end{figure}

In Figure~\ref{fig:uni} we present a sample election and results of
choosing a committee of size $34$ (committee members are depicted as
large red dots).  At first sight, we see that SNTV, STV,
Chamberlin--Courant, and Greedy-Monroe do a good job in terms of
representing the smaller centrist population. On the other hand, Random-District-FPP and
Random-District Borda seem to provide very scattered, erratic results,
with FPP underrepresenting the minority, and Random-District Borda
overrepresenting it. Bloc ignores the minority completely, whereas
$k$-Borda seems to focus on it completely.

\begin{figure}
  \centering
  \includegraphics[height=8cm]{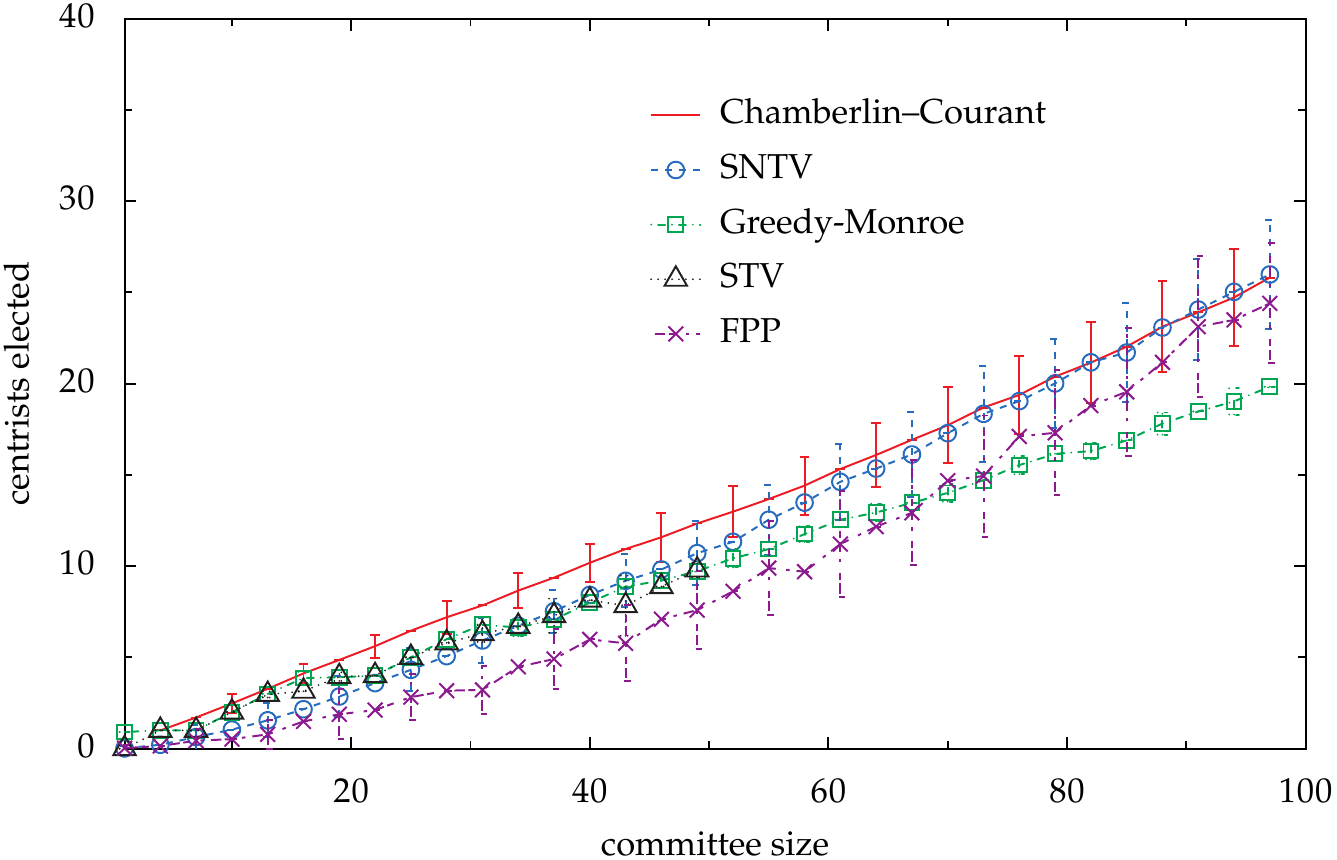}
  \caption{\label{fig:partyC-uni}Average number of candidates from the
    centrist party selected by SNTV, STV, Chamberlin--Courant,
    Greedy-Monroe, and FPP. Vertical bars indicate standard
    deviation.}
\end{figure}

\paragraph{Proportionality}  A key concept in the theory of
representation is the concept of proportionality.  This notion has a
clear meaning when votes and candidates are labeled alike: When voters
vote for parties, one can check whether the number of elected
candidates from a party is proportional to the party's score. When
delegates are elected by districts, one can check whether or not the
number of seats allocated to each district is proportional to the
population of the district.%

In order to check if our four election rules that did best in terms of voter
representation indeed represent the centrist group proportionally, we
can think of the candidates as belonging themselves to the three
groups. We simply consider that a candidate ``belongs'' to the group
closest to her location.

We have generated 65 elections according to the above-described
scheme; for each, we have computed committees of size 1 to 97 (with a
step of 3), and computed how many candidates from each party were
selected.
We show the results in Figure~\ref{fig:partyC-uni} (we also include
Random-District-FPP for comparison).  We see that, after all, there is
some difference between the proportionality achieved by our four
rules. While STV and Greedy-Monroe seem to select roughly 20\% of the
candidates from the centrist party (the desired number), SNTV and
Chamberlin--Courant overshoot. Greedy-Monroe does even better than STV
because it is far more stable (the standard deviation of the results
for Greedy-Monroe is noticeably smaller than for STV). FPP undershoots
significantly.

\begin{figure}[p!]
    \centering
    \begin{subfigure}{7cm}
      \centering
      \caption{\label{fig:kBorda}Results for SNTV}
      \includegraphics[height=3.25cm]{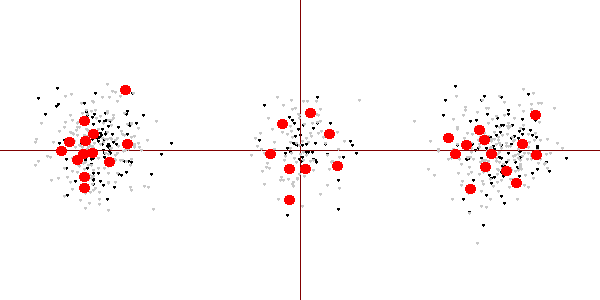}  
    \end{subfigure}
    \begin{subfigure}{7cm}
      \centering
      \caption{\label{fig:Bloc}Results for STV}
      \includegraphics[height=3.25cm]{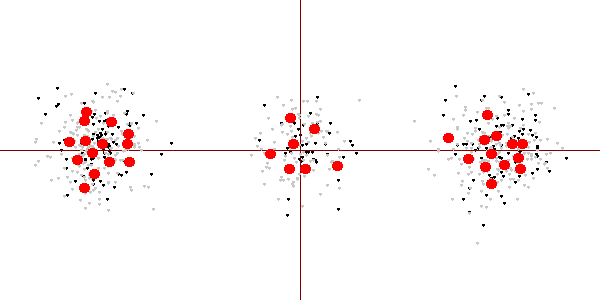}
    \end{subfigure}

    \bigskip\bigskip

    \begin{subfigure}{7cm}
      \centering
      \caption{\label{fig:kBorda}Results for Chamberlin--Courant}
      \includegraphics[height=3.25cm]{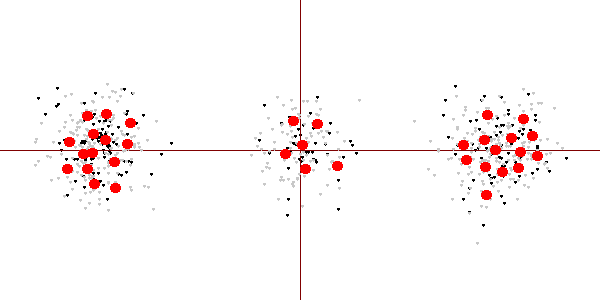}  
    \end{subfigure}
    \begin{subfigure}{7cm}
      \centering
      \caption{\label{fig:Bloc}Results for Greedy-Monroe}
      \includegraphics[height=3.25cm]{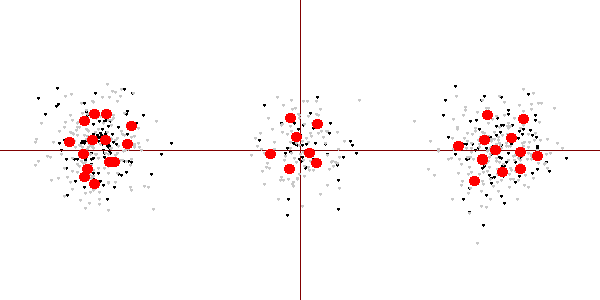}
    \end{subfigure}

    \bigskip\bigskip

    \begin{subfigure}{7cm}
      \centering
      \caption{\label{fig:kBorda}Results for FPP}
      \includegraphics[height=3.25cm]{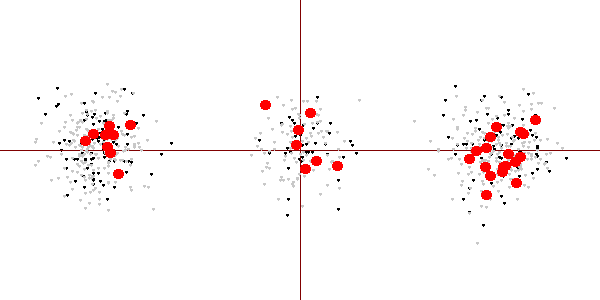}  
    \end{subfigure}
    \begin{subfigure}{7cm}
      \centering
      \caption{\label{fig:Bloc}Results for District-Based Borda}
      \includegraphics[height=3.25cm]{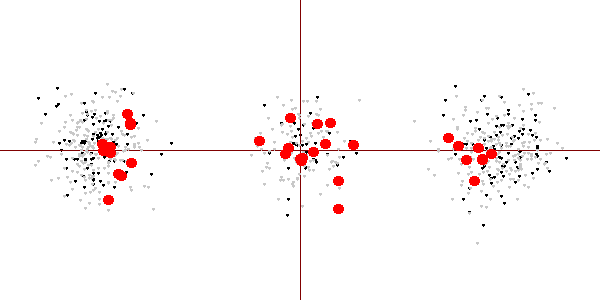}
    \end{subfigure}

    \bigskip\bigskip

    \begin{subfigure}{7cm}
      \centering
      \caption{\label{fig:kBorda}Results for Borda}
      \includegraphics[height=3.25cm]{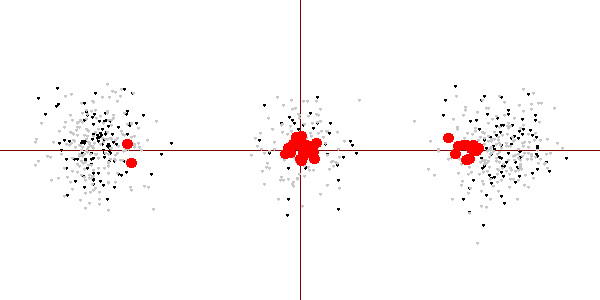}  
    \end{subfigure}
    \begin{subfigure}{7cm}
      \centering
      \caption{\label{fig:Bloc}Results for Bloc}
      \includegraphics[height=3.25cm]{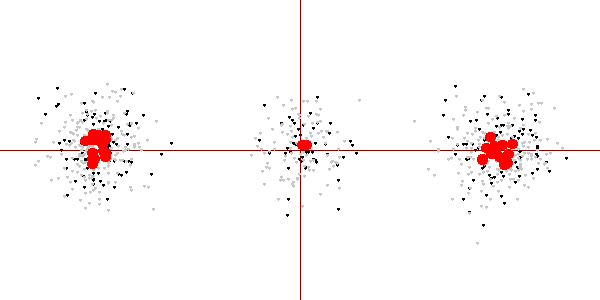}
    \end{subfigure}

    \caption{\label{fig:gau}Results for two big parties and a smaller
      centrist party, for committee size $k=34$, for the case where
      candidates and voters follow the same distribution.}
\end{figure}

To verify the robustness of our results with respect to the location
of the candidates, we have repeated our experiment for the same
distribution of voters (however, we have now used 500 voters instead
of 250) and for 250 candidates distributed in the same way as the
voters. That is, now we assumed that the structure of preferences that
lead to the formation of the groups is also present among the
candidates.  This is the same ``citizen-candidate'' hypothesis that
was made in the first experiment, and it gives a more direct way of
modeling party affiliations of candidates.  In Figure~\ref{fig:gau} we
present the results for a sample election, for committee size $k =
34$.  Comparing to Figure~\ref{fig:uni}, we can see that now all the
rules seem to behave more proportionally. We believe that the reason
for this fact is that, in some sense, the rules have far fewer
candidates to choose from; there are no maverick candidates all over
the political spectrum that would distract the voters. However, still
it is visible that our four proportional representation rules seem to
be doing best, that $k$-Borda overrepresents the center, and that Bloc
underrepresents it.  Interestingly, district-based rules seem to be
doing fine.

\begin{figure}
  \centering
  \includegraphics[height=8cm]{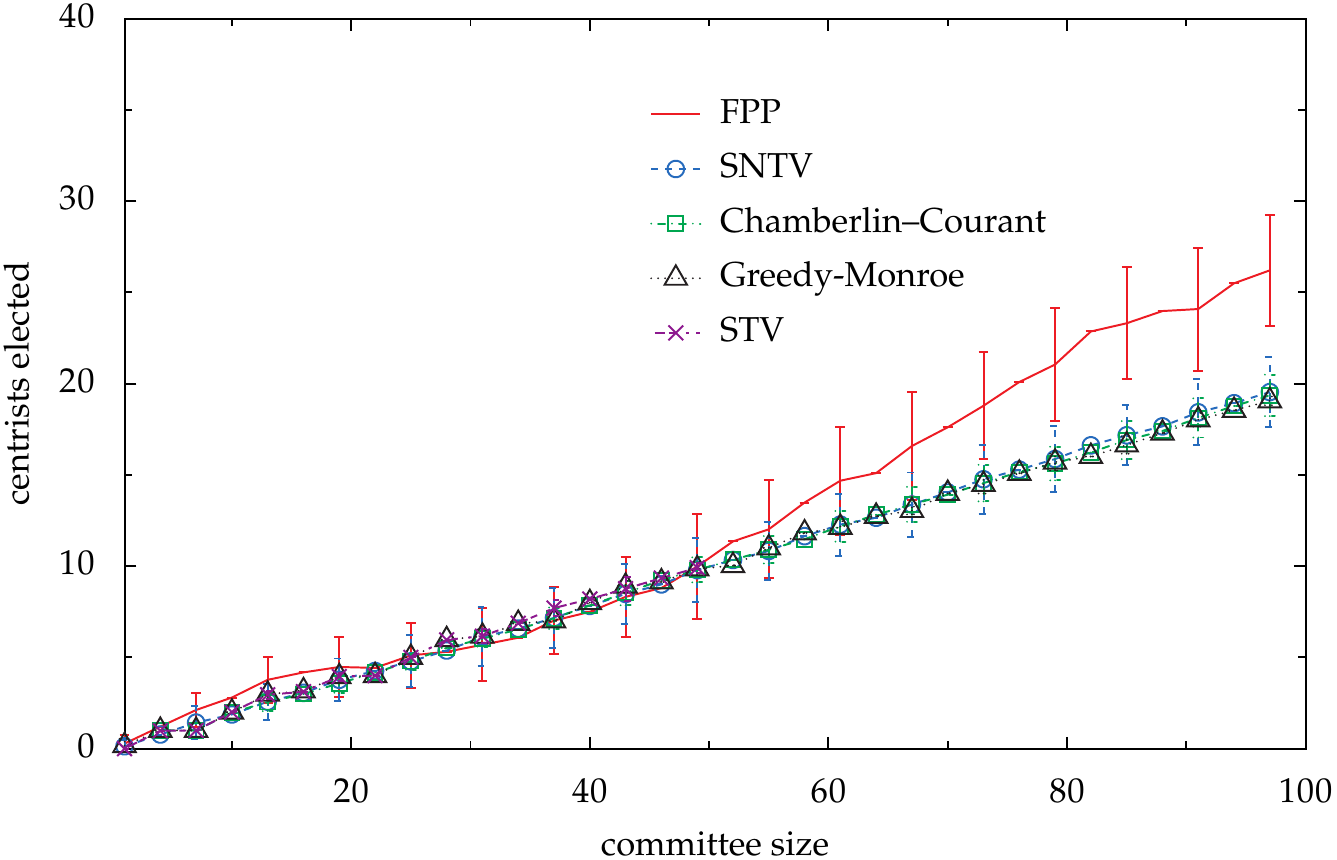}
  \caption{\label{fig:partyC-gau}Average number of candidates from the
    centrist party selected by SNTV, STV, Chamberlin--Courant,
    Greedy-Monroe, and FPP. Vertical bars indicated standard
    deviation.}
\end{figure}

In Figure~\ref{fig:partyC-gau} we show the average number of
candidates from the centrist party elected by the four rules (and
Random-Districts-FPP; added for comparison; this is a result from generating
100 elections). As one might have
expected from Figure~\ref{fig:gau}, the scenario where candidates and
voters are identically distributed is easy for the rules that aim at
proportional representation by design. All these rules perform
well. Interestingly, for larger committees FPP overshoots
significantly.

\section{Conclusion and further work}
Firstly, we have confirmed that the choice of a voting rule has a
profound effect on representation of minorities and any electoral
system designer must take this into consideration.  Secondly, we have
initiated the study on evaluation of multiwinner voting rules with
respect to their ability to provide faithfully represent the voters. To
this end, we have considered two parameters: (1) the average
misrepresentation, and (2) the proportion of voters elected from a
smaller centrist party.  

It turned out that among our rules, STV,
SNTV, Chamberlin--Courant, and Greedy-Monroe, four rules that to large
extent were designed to achieve proportional representation, indeed
achieve it. Nonetheless, we have seen that additional mechanisms for
ensuring proportionality built into Greedy-Monroe (and, to some
degree, into STV) indeed give them advantage in more challenging
settings. On the other hand, rules based on random-districting (in
particular FPP) turn out to be not reliable. Naturally, rules that
were designed with other principles in mind than proportional
representation ($k$-Borda and Bloc, in our case) do not fare well
compared to the others.  Since we located the minority in the center
of the political spectrum, we cannot say, at this point, if the same
results would hold in other cases.  We consider this work only a
starting point and we are working on further experiments.  \medskip

\bibliographystyle{abbrv}
\bibliography{grypiotr2006}  %
\end{document}